\documentclass[aps,prbrc,twocolumn,floatfix]{revtex4}

\usepackage{graphicx}

\begin{document}


\title{Ultrahigh field electron cyclotron
resonance absorption in In$_{1-x}$Mn$_x$As films}


\author{M. A. Zudov}
\thanks{Present address: Physics Department, University of Utah,
Salt Lake City, Utah 84112, U.S.A.}
\author{J. Kono}
\thanks{To whom correspondence should be addressed.}
\homepage{http://www.ece.rice.edu/~kono}
\email{kono@rice.edu}
\affiliation{Department of Electrical and Computer Engineering,
Rice Quantum Institute, and\\
Center for Nanoscale Science and Technology,
Rice University, Houston, Texas 77005, U.S.A.}
\author{Y. H. Matsuda}
\thanks{Present address: Department of Physics, Faculty of Science,
Okayama University, Okayama, Japan.}
\author{T. Ikaida}
\author{N. Miura}
\affiliation{Institute for Solid State Physics, University of Tokyo,
Kashiwanoha, Kashiwa, Chiba 277-8581, Japan}
\author{H. Munekata}
\affiliation{Imaging Science and Engineering Laboratory, Tokyo
Institute of Technology, Yokohama, Kanagawa 226-8503,
Japan}
\author{G. D. Sanders}
\author{Y. Sun}
\author{C. J. Stanton}
\affiliation{Department of Physics, University of Florida,
Gainesville, Florida 32611, U.S.A.}

\date{\today}
\begin{abstract}

We have carried out an ultrahigh field cyclotron resonance
study of $n$-type In$_{1-x}$Mn$_x$As films, with Mn
composition $x$ ranging from 0 to 12\%, grown on GaAs by low
temperature molecular beam epitaxy.  We observe that the electron
cyclotron resonance peak shifts to lower field  with increasing $x$.
A detailed comparison of
experimental results with calculations based on a modified
Pidgeon-Brown model allows us to estimate the {\em s-d} and {\em p-d}
exchange coupling constants, $\alpha$ and $\beta$, for this important
III-V dilute magnetic semiconductor system.

\end{abstract}
\maketitle




InMnAs alloys and their heterostructures with AlGaSb, the first grown
III-V dilute magnetic semiconductor (DMS)
\cite{munekata,ohno,munekata1}, serve as a prototype for implementing
electron and hole spin degrees of freedom in semiconductors.  Recent
experiments have demonstrated the feasibility of controlling
ferromagnetism in these systems optically \cite{koshihara} and
electrically \cite{ohno1}.  Understanding their electronic, transport,
and optical properties is crucial for designing novel ferromagnetic
semiconductor devices with high Curie temperatures.  However, their
basic band parameters such as effective masses and $g$-factors have not
been accurately determined.

The localized Mn spins strongly influence the delocalized conduction
and valence band states through the {\em s-d} and {\em p-d} exchange
interactions.  These interactions are usually parameterized as $\alpha$
and $\beta$, respectively \cite{furdyna}.  Determining these parameters
is important for understanding the nature of Mn electron states and
their mixing with delocalized carrier states.  In narrow gap
semiconductors like InMnAs, due to strong interband mixing, $\alpha$
and $\beta$ can influence {\it both} the conduction and valence bands.
This is in contrast to wide gap semiconductors where $\alpha$
influences primarily the conduction band and $\beta$ the valence band.
In addition, in narrow gap semiconductors, due to the strong interband
mixing, the coupling to the Mn spins does not effect all Landau levels
by the same amount.  As a result, the electron cyclotron resonance (CR)
peak can shift as a function of $x$, the Mn concentration. This can be
a sensitive method for estimating these exchange parameters.
A recent CR study on Cd$_{1-x}$Mn$_x$Te \cite{matsuda2}
showed that the electron mass is strongly affected by
$sp-d$ hybridization.  To our knowledge, however, there have been no
CR studies on any III-V DMS
systems (for our preliminary results, see \cite{matsuda,matsuda1}).

In this paper we describe the dependence of electron CR observed in
$n$-type In$_{1-x}$Mn$_x$As films on $x$.  A
detailed comparison with theoretical calculations provides insight
into the effects of Mn ions on the Landau and Zeeman splittings of the
conduction band states.

\begin{table}[bth]
\caption{Densities, mobilities, and cyclotron masses for the four
samples studied.  The densities are in units of cm$^{-3}$ and the
mobilities cm$^2$/Vs.  The masses were obtained at a
photon energy of 117 meV (or $\lambda$ = 10.6 $\mu$m).}
\label{table1}
\begin{ruledtabular}
\begin{tabular*}{\hsize}{l@{\extracolsep{0ptplus1fil}}c@{\extracolsep{
0ptplus1fil}}c@{\extracolsep{0ptplus1fil}}c@{\extracolsep{0ptplus1fil}}c}
Mn content $x$ & 0 & 0.025 & 0.050 & 0.120 \\
\colrule
Density (4.2 K) & $\sim$1.0$\times$10$^{17}$ & 1.0$\times$10$^{16}$ &
0.9$\times$10$^{16}$ & 1.0$\times$10$^{16}$ \\
Density (290 K) & $\sim$1.0$\times$10$^{17}$ & 2.1$\times$10$^{17}$ &
1.8$\times$10$^{17}$ & 7.0$\times$10$^{16}$ \\
Mobility (4.2 K) & $\sim$4000 & 1300 & 1200 & 450 \\
Mobility (290 K) & $\sim$4000 & 400 & 375 & 450 \\
$m$/$m_0$ (30 K) & 0.0342 & 0.0303 & 0.0274 & 0.0263 \\
$m$/$m_0$ (290 K) & 0.0341 & 0.0334 & 0.0325 & 0.0272 \\
\end{tabular*}
\end{ruledtabular}
\end{table}

We studied four $\sim$2-$\mu$m-thick In$_{1-x}$Mn$_x$As films with $x$
= 0, 0.025, 0.050 and 0.120 by ultrahigh-field magneto-absorption
spectroscopy.  The films were grown by molecular beam epitaxy on
semi-insulating GaAs substrates at 200$^{\circ}$C.  All the samples
were $n$-type and did not show ferromagnetism down to 1.5 K.  The
electron densities and mobilities deduced from Hall measurements are
listed in Table \ref{table1}.  The single-turn coil technique
\cite{nakao} was used to generate ultrahigh magnetic fields with pulse
duration $\sim$7 $\mu$s.  The sample and pick-up coil were placed in a
helium flow cryostat.
We used the 10.6 $\mu$m line
from a CO$_2$ laser and produced circular polarization using a CdS
quarter-wave plate.  The transmitted radiation was detected by a
fast HgCdTe photovoltaic detector.
\begin{figure}
\includegraphics[scale=0.85]{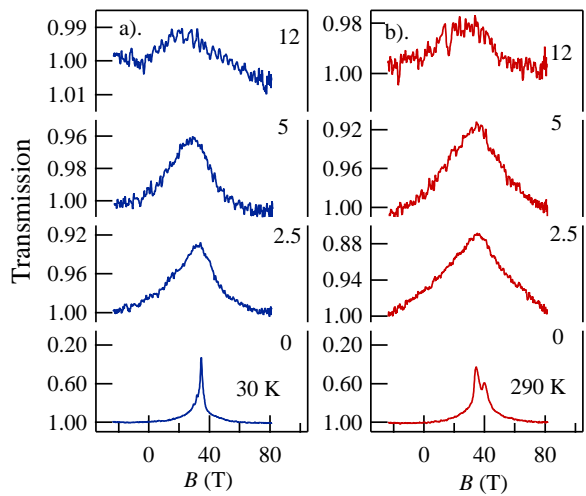}
\caption{Experimental CR spectra for different Mn contents
taken for 290 (a) and 30 K (b).  The wavelength of the laser is
fixed at 10.6 $\mu$m with electron-active circular polarization
and the field $B$ is swept.
The resonance position shifts to lower $B$ with increasing $x$.
The $x$ values are 0\%, 2.5\%, 5\% and 12\%.}
\label{figcr}
\end{figure}

\begin{figure}
\begin{center}
\includegraphics[scale=0.40]{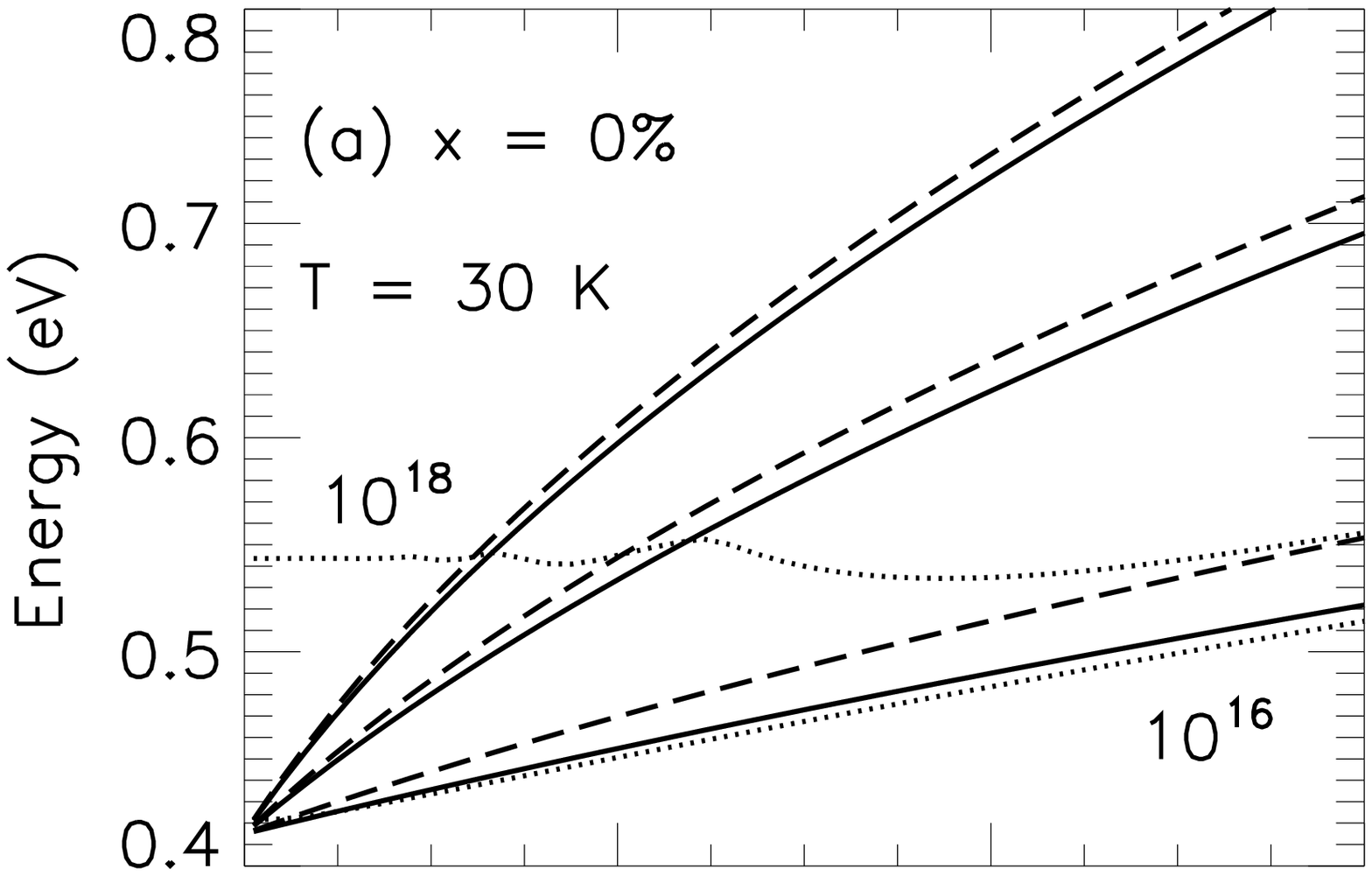}
\vskip -0.1in
\includegraphics[scale=0.40]{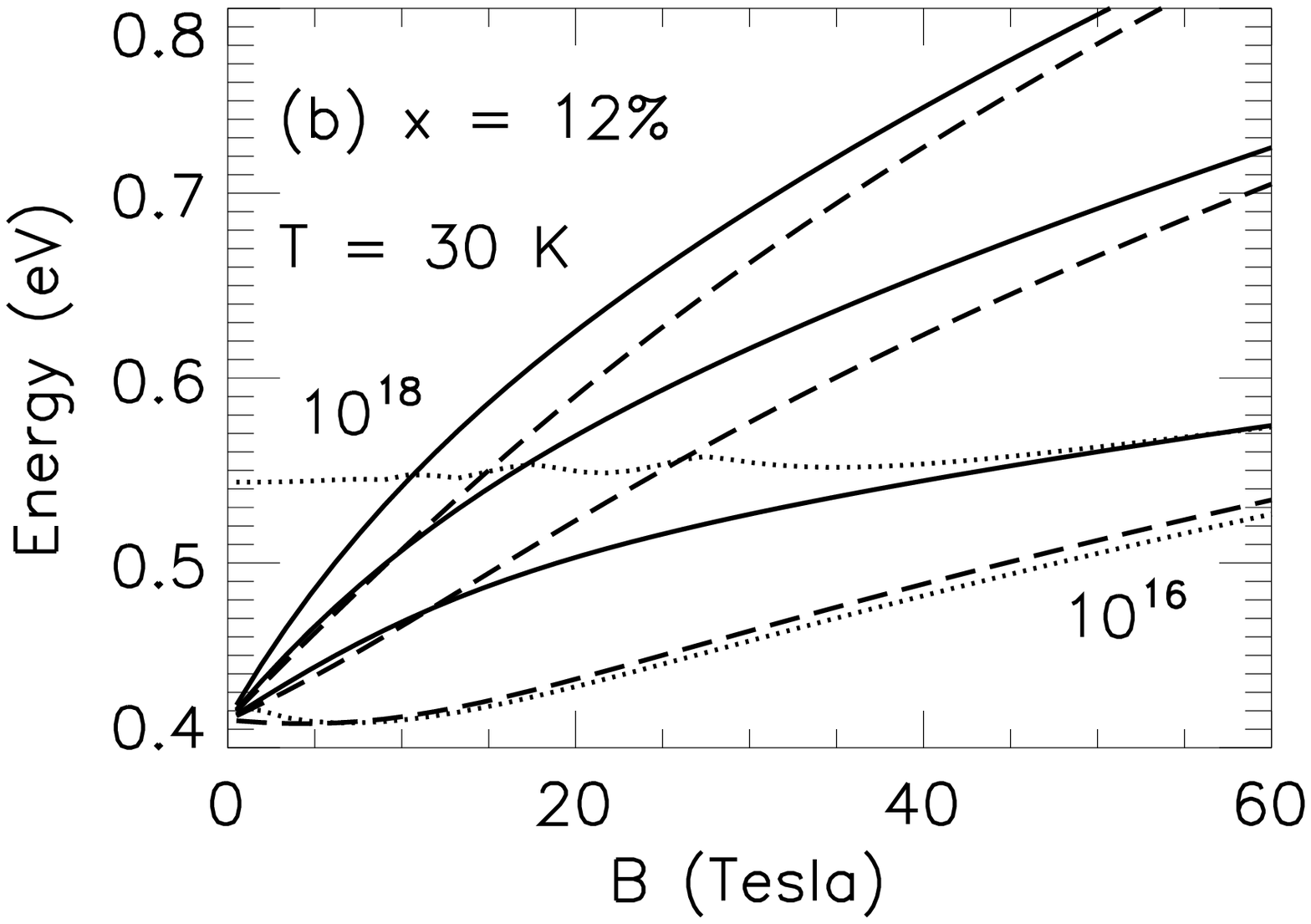}
\end{center}
\caption {Electron Landau levels as a function of magnetic field, $B$,
at 30 K for (a) $x = 0\%$ and (b) $x = 12\%$. Spin up (down) levels are
shown as solid (dashed) lines.
The Fermi levels are also indicated for electron concentrations
of $10^{16}$ and $10^{18}$ cm$^{-3}$. }
\label{theory3}
\end{figure}

Typical measured CR spectra at 30 K and 290 K are
shown in Figs. \ref{figcr} (a) and (b), respectively.  Note that to
compare the transmission with absorption calculations, the transmission
increases in the negative $y$ direction.  Each figure shows spectra for
all four samples labeled by the corresponding Mn compositions from 0 to
12 \%.  All the samples show pronounced absorption peaks (or
transmission dips) and the resonance field {\em decreases} with
increasing $x$.  Increasing $x$ from 0 to 12\% results in a $\sim$25 \%
decrease in cyclotron mass (see Table \ref{table1}).  It is important
to note that at resonance, the densities and fields are such that only
the lowest Landau level for each spin type is occupied (see Fig.
\ref{theory3}).  Thus, all the electrons were in the lowest Landau
level for a given spin even at room temperature, precluding any
density-dependent mass due to nonparabolicity (expected at zero or low
magnetic fields) as the cause of the observed trend.

At high temperatures (e.g., Fig. \ref{figcr}(b)) the $x$ = 0 sample
clearly shows nonparabolicity-induced CR spin-splitting with the weaker
(stronger) peak originating from the lowest spin-down (spin-up) Landau
level, while the other three samples do not show such splitting.  The
reason for the absence of splitting in the Mn-doped samples is a
combination of 1) their low mobilities (which lead to substantial
broadening) and 2) the large effective $g$-factors due to the Mn
ions; especially in samples with large $x$ only the spin-down
level is substantially thermally populated (cf. Fig. \ref{theory3}).

We also performed midinfrared interband absorption measurements
at various temperatures using Fourier-transform infrared
spectroscopy, which revealed no significant
$x$-dependence of the band gap except for the different amounts of
Burstein-Moss shifts due to the different carrier densities.

We calculated the conduction band Landau levels based on a modified
Pidgeon-Brown model \cite{pidgeon,kossut} including full $k_z$
dependence and also {\it sp-d} exchange coupling of the electrons and
holes to the Mn ions.  This is an 8 $\times$ 8 band ${\bf k}\cdot{\bf
p}$ method with a magnetic field along
the [001] direction.  We used a standard set of band parameters for
InAs \cite{meyer}, neglecting any $x$-dependence of these parameters.
We vary the gap with $T$ according to $E_g(T)= 0.417  -
0.000276 T^2 /(93 +T) {\rm eV}$ \cite{meyer}.  As seen in Table I, the
mass does not vary significantly with $T$.  Other low field
CR studies of InAs \cite{Palik} show a very weak
$T$-dependence to the cyclotron mass and also take into account
polaron effects.  Since we are mainly interested in the $x$ dependence
of the mass, for simplicity, we adjust the $F$ parameter to keep the
mass constant with $T$.  Alternatively, one might wish to keep
$F$ constant and vary the ${\bf k}\cdot{\bf p}$ optical matrix element
$E_p$.
\begin{figure}
\begin{center}
\includegraphics[scale=0.5]{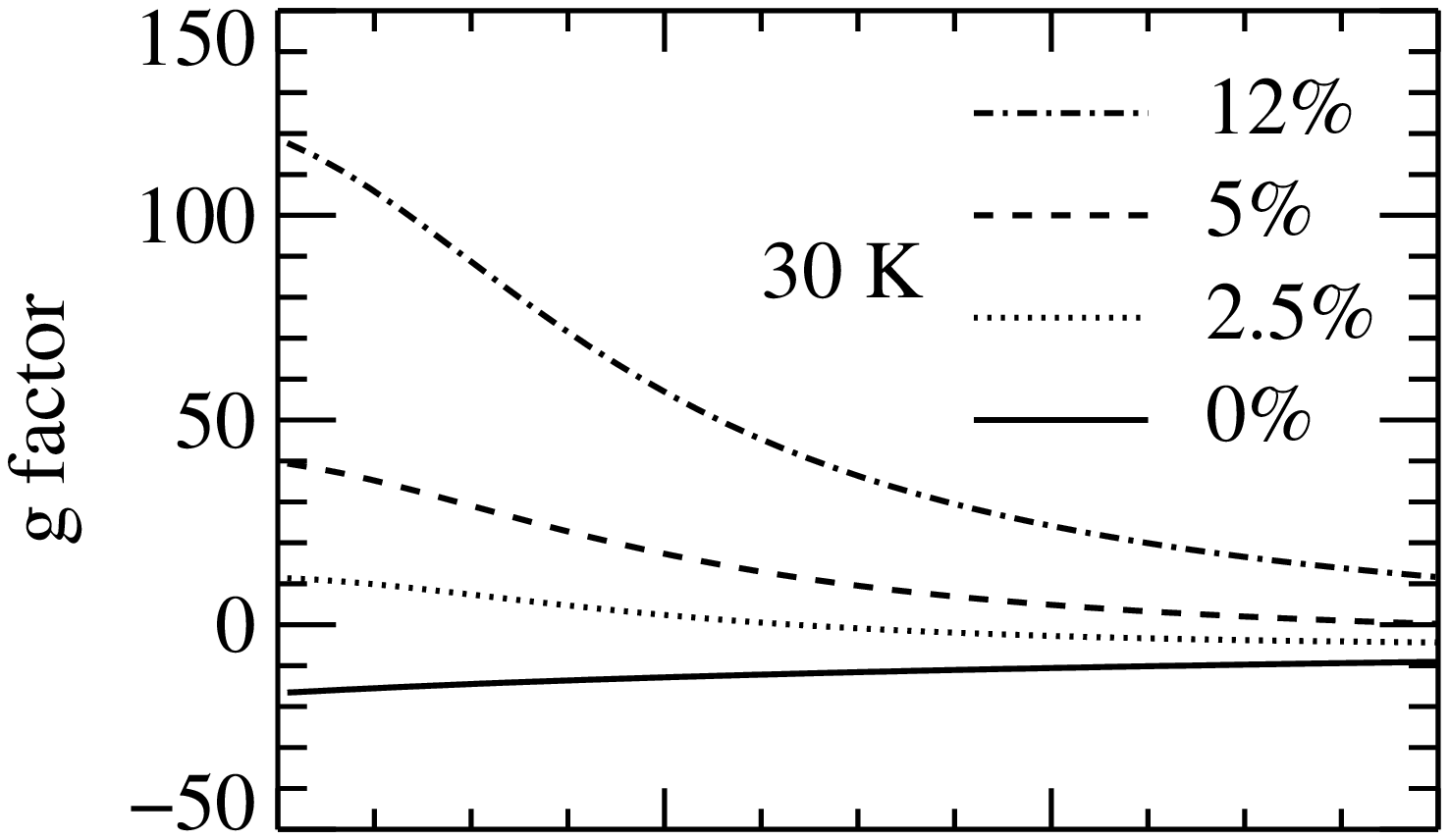}
\vskip -0.65in
\includegraphics[scale=0.5]{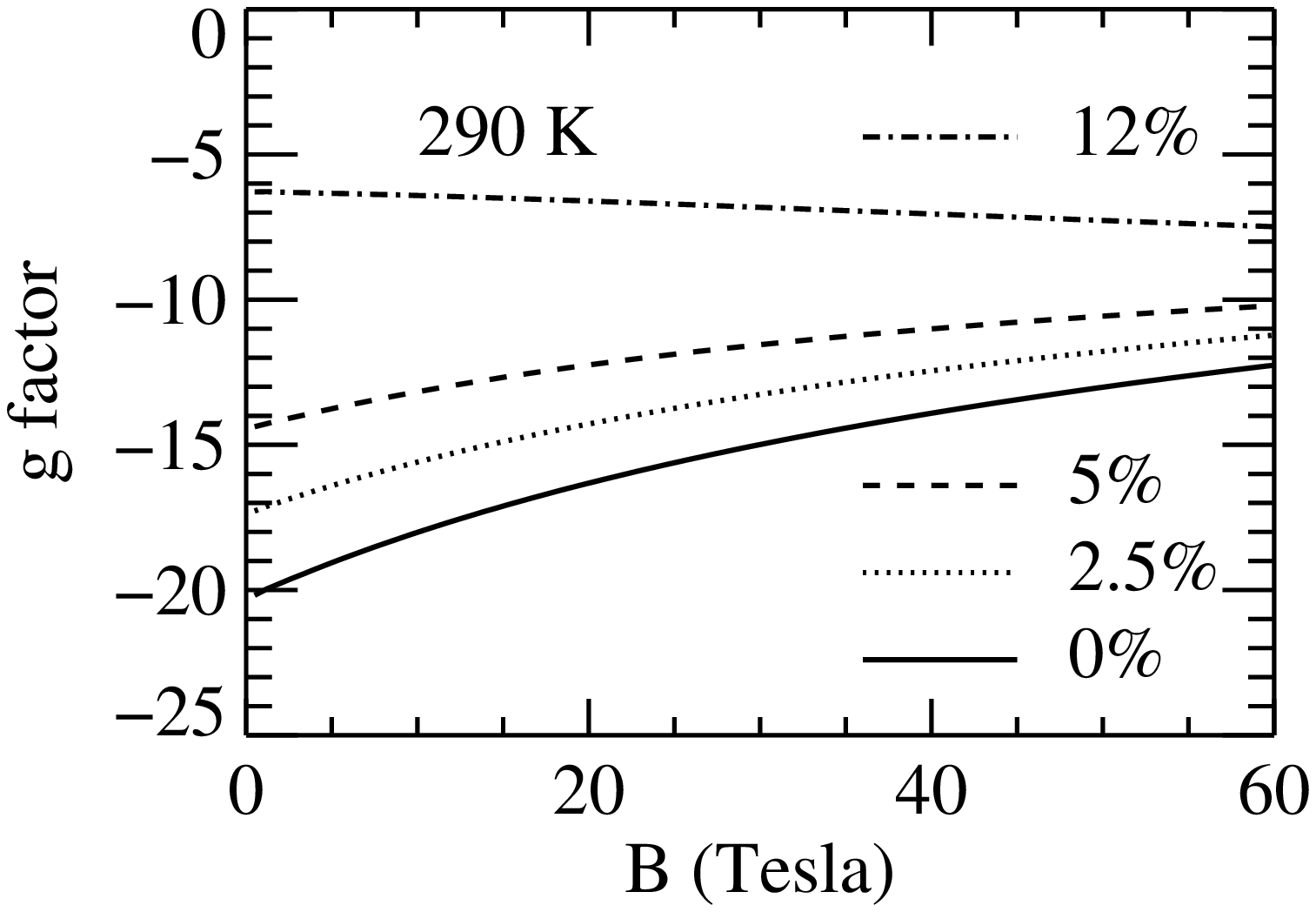}
\end{center}
\caption {Electron effective $g$ factors for $In_{1-x}Mn_{x}As$
as a function of applied magnetic field for several values of
the $Mn$ concentration, $x$.  The upper and lower panels
correspond to $T=30\ \mbox{K}$ and $T=290\ \mbox{K}$, respectively.
}
\label{gtheory}
\end{figure}

Since InMnAs is a narrow gap semiconductor, the coupling between the
valence and conduction bands is strong, and thus, not only $\alpha$ but
also $\beta$ is important in calculating the conduction band Landau
levels.  A photoemission experiment \cite{okabayashi} reports
$\beta$ = $-$0.7 eV, and a theoretical estimate
\cite{dietl} suggests that $\beta \approx -0.98$ eV.  No value
has been reported for $\alpha$, to our knowledge.

We chose $\beta$ = $-$1.0 eV and $\alpha$ = 0.5 eV as the values which
best represent the observed trends though these values can change
depending upon the other parameters used in the ${\bf k}\cdot{\bf p} $
calculation.  Figures \ref{theory3}(a) and \ref{theory3}(b) show the
lowest three Landau levels (spin up and down) in the conduction band
for the $x$ = 0 and 12\% samples.  In addition, the Fermi levels are
calculated and plotted (dotted lines) for carrier densities of $10^{18}
{\rm cm}^{-3}$ and $10^{16} {\rm cm}^{-3}$.
Both the mass and $g$-factor are
strongly energy and magnetic field dependent as well as $x$ dependent.
We can also see that in the $x$ = 12\% sample the {\em spin order is
reversed} and the spin splitting is significantly enhanced by the
presence of Mn ions; the latter explains the absence of CR spin
splitting in the 12\% sample.  We found that both the sign and values
of $\alpha$ and $\beta$ are critically important in explaining the
experimental data and therefore it is a good method for determining
these parameters.  In various investigations of II-VI DMS's it has been
established that $\alpha$ and $\beta$ differ in sign and usually the
absolute value of $\beta$ is greater than $\alpha$
\cite{furdyna,kossut}, which is consistent with our results.

In Fig. \ref{gtheory}, we show the effective $g$-factors for electrons
in the lowest Landau level in In$_{1-x}$Mn$_x$As
as a function of applied magnetic field and Mn doping concentration,
$x$.  The curves are shown for temperatures of 30 K and 290 K,
respectively.  As can be seen, at 30 K and high Mn concentration, the
effective $g$-factor is extremely large ($>100$) and has the opposite sign
of the undoped sample.  Due to Mn doping and the large nonparabolicity
of the narrow gap semiconductor, the $g$-factor is very dependent on magnetic field and
is reduced substantially at high magnetic fields and temperatures.

Based on the calculated Landau levels,  we also calculated the cyclotron
resonance spectra for the four  samples at 30 K and 290 K.
The magneto-optical absorption coefficient at the photon
energy $\hbar \omega$ is
\cite{Bassani}
\begin{equation}
\alpha(\hbar \omega)=
\frac{\hbar \omega}{(\hbar c) n_r}\ \epsilon_2(\hbar \omega)
\end{equation}
where $\epsilon_2(\hbar \omega)$ is the imaginary part of the dielectric
function and $n_r$ is the index of refraction.
The imaginary part of the dielectric function is found using Fermi's
golden rule. The result is

\begin{eqnarray}
&&
\epsilon_2(\hbar \omega) =  \frac{e^2}{\lambda^2(\hbar \omega)^2}
\sum_{n,\nu; n',\nu'} \int_{-\infty}^{\infty} dk_z \
\arrowvert \hat{e} \cdot \vec{P}_{n,\nu}^{n',\nu'}(k_z) \arrowvert^2
\nonumber \\
&&
\times \left( f_{n,\nu}(k_z) - f_{n',\nu'}(k_z) \right)
\delta
\left( \Delta E_{n',\nu'}^{n,\nu}(k_z) - \hbar \omega \right),
\label{epsilon2}
\end{eqnarray}
where
$\Delta E_{n',\nu'}^{n,\nu}(k_z)=E_{n',\nu'}(k_z)-E_{n,\nu}(k_z)$
is the transition energy and $\lambda$ the magnetic length.
The function $f_{n,\nu}(k_z)$ in Eq. (\ref{epsilon2})
is the probability that the state
$(n,\nu,k_z)$, with energy $E_{n,\nu}(k_z)$, is occupied. It is
given by the Fermi distribution function and depends on the net carrier
density.

\begin{figure}
\includegraphics[scale=0.6]{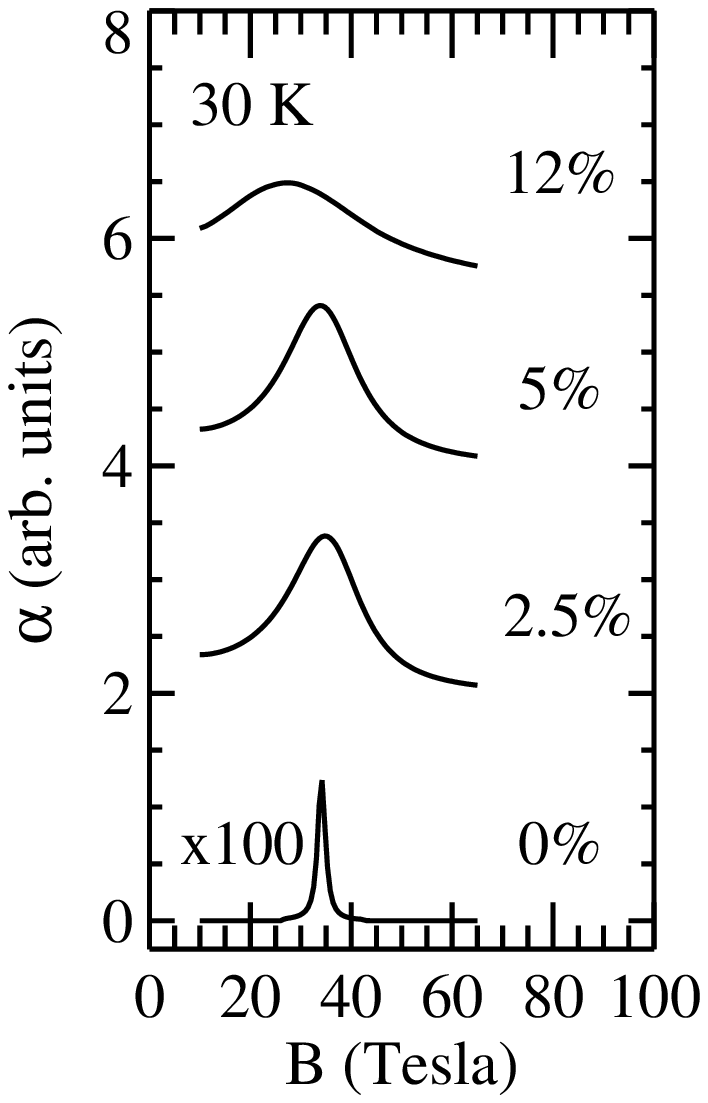}%
\includegraphics[scale=0.6]{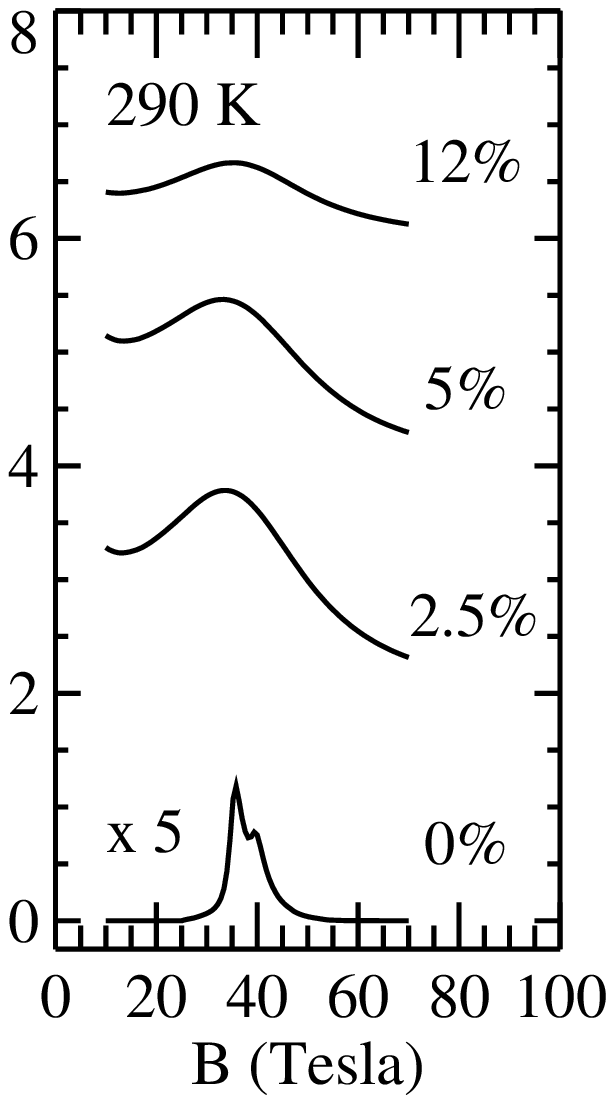}
\caption {Calculated CR absorption
as a function of magnetic field
at 30 K (a) and 290 K (b).
The curves are calculated based on the Pidgeon-Brown model
and the golden rule for absorption. The curves are broadened based
on the mobilities reported in Table I.}
\label{theory1}
\end{figure}

Figures \ref{theory1}(a) and \ref{theory1}(b) show the calculated
CR absorption coefficient for electron-active
circularly polarized 10.6 $\mu$m light in the Faraday configuration as
a function of magnetic field at 30 K and 290 K, respectively. Densities
for each sample are given in Table I.  In the calculation, the curves
were broadened based on the mobilities of the samples.  The broadening
used was: ($T$ = 30 K, 0\% - 4 meV, 2.5\% - 40 meV, 5\% - 40 meV and 12\%
- 80 meV); ($T$ = 290 K, 0\% - 4 meV, 2.5\% - 80 meV, 5\% - 80 meV, 12\%
- 80 meV).

At $T$ = 30 K, we see a shift in the CR peak as a
function of doping in agreement with Fig. \ref{figcr} (a).   For
$T$ = 290 K, we see the presence of two peaks in the pure InAs sample.
The second peak originates from the thermal population of the lowest
spin-down Landau level (cf. Fig. \ref{theory3}).  The peak does not
shift as much with doping as it did at low temperature.  This results
from the temperature dependence of the average Mn spin.
We believe that the Brillouin function used for calculating the average
Mn spin becomes inadequate at large $x$ and/or high
temperature due to its neglect of Mn-Mn interactions such as
pairing and clustering.

In summary, we have observed electron CR in
In$_{1-x}$Mn$_x$As for the first time
and studied its dependence on the Mn concentration, $x$.
Owing to band mixing in this narrow gap system, the
CR peak actually shifts with $x$; we observed a $\sim$25\%
mass reduction by increasing $x$ from 0 to 0.12.
Theoretical calculations
based on an $8 \times 8$ band {\bf k}$\cdot${\bf p} model
successfully reproduce this behavior.
These results demonstrate that CR can be used for
determining the $sp-d$, electron/hole - $Mn$ ion exchange
interactions $\alpha$ and $\beta$.
We obtained $\beta$ = $-$1.0 eV and $\alpha$ = 0.5 eV as the values
which best represent the observed trends.  We also showed that
at low temperatures and high $x$ the effective $g$-factor is
extremely large ($>100$) and has the opposite sign of the
$x$ = 0 sample; it is very dependent on magnetic field and
reduced substantially at high magnetic fields and temperatures.
These findings should be useful for designing novel spin-based
semiconductor devices.

This work was supported by DARPA through grant No. MDA972-00-1-0034 and
the NEDO International Joint Research Program.

\end{document}